\documentstyle[11pt,epsfig]{article}
\title{\bf Effect of RGUP on the nonlinear Klein-Gordon model with spontaneous symmetry breaking}

\author{S. Miraboutalebi$^1$\footnote{~s$_{-}$mirabotalebi@iau-tnb.ac.ir}, F. Ahmadi$^2$ and A. Jahangiri$^1$\\
$^1${\small Department of Physics, Islamic Azad University, North Tehran Branch, Tehran, 1651153311, Iran.}\\
$^2${\small Department of Physics, Shahid Rajaee Teacher Training University, Tehran 1678815811, Iran.}\\}
\begin{document}
\maketitle
\begin{abstract}

In the framework of the relativistic generalized uncertainty principle (RGUP), we study the nonlinear Klein-Gordon field with $\phi^4$ self-interaction. This generalization comes from the quantum gravity theories that predict the existence of a minimal measurable length scale. The quantum gravity effects slightly modify the momentum and change the Lagrangian density of the field. This model, specifically in the context of spontaneous symmetry breaking, is so fruitful and leads us to many interesting phenomena. Due to the complexity of the consequent equations, the model is normally studied via perturbation mechanisms. However, here, we apply a generalized $tanh$-method which supposed that the field solutions should be $sech$-functions of segments of space and time. By this extended method, we find some more general solutions with respect to our previous work \cite{Mir1}. Several solutions can be found but we select only the normalizable and bounded solitary fields. The energy spectrum of each field solution is obtained in an analytical form and discussed. The modification parameter of the corresponding RGUP is estimated by using the rest energy of the mass of the Higgs boson.

\end{abstract}



\section{Introduction}

Almost every physical phenomenon in Universe can be described by theories of Quantum Mechanics (QM) and General Relativity (GR). These theories contain the fundamental models that describe matter and its essential interactions and also predict many experimentally observable results. Perhaps the most important application of QM is when it is applied to Classical Field Theory (CFT) to formulate Quantum Field Theory (QFT). The most prominent model prepared with the use of QFT is the Standard Model of particle physics. The Standard Model, fortunately, describes well the behavior of the elementary particles under the three fundamental interactions of nature, namely electromagnetism, and weak and strong nuclear forces. In addition, this theory succeeds in predicting the way in which leptons and weak bosons acquire mass, and also it predicts the discovery of a plethora of particles, like the Higgs boson \cite{Engl, Higg, Gura}.

Despite many triumphs of QM and its extensions, such as the Standard Model, it cannot be considered an ultimate general theory describing the world. Because they are incapable of explaining the behavior of the elementary particles under the influence of gravitational interactions. QM considers space-time as a flat stationary background, while GR considers it a dynamic object which interacts with matter and energy. On the other hand, GR can not formulate the probabilistic nature of the observables described by QM. Therefore, one needs to construct a new theory that will give rise to QM and GR in their respective domains . Based on previous experience, one hopes that a physical theory that contains both QM and GR and describes all fundamental interactions of the matter shall be a QFT.

GR describes the gravitational interaction as a CFT with the Einstein-Hilbert action \cite{Eins, AEins}. It should be noticed that, unlike the other QFTs, the coupling constant of GR has a dimension that results from Newton's constants $G$. The dimensionful coupling constant gives rise to a length scale and means that when one tries to quantize GR using the standard methods defined in the framework of QFT \cite{Kaku}, one runs into difficulties. In particular, experimentally measurable quantities such as scattering cross-sections of processes involving gravity \cite{Nesv1, Nesv2}, expectedly turn out to be infinite. The infinite nature of such results is common and is usually treated by renormalization procedures\cite{Frad1, Frad2}. The application of renormalization processes to perturbatively quantized gravity, on the other hand, fail to get a finite measurable answer. The approaches and theories tackling the problem of quantizing gravity have come to be known as theories of Quantum Gravity (QG). Some examples of such theories include String Theory (ST), Loop Quantum Gravity (LQG), and Doubly Special Relativity (DSR) \cite{Jalalzadeh}. All of these approaches, and many others, have been somewhat successful. One common aspect of all these models of QG, as explained in \cite{Luis, Hossen, vas}, is that they all indicate the existence of a minimum measurable length.

In the presence of this minimal length, the Heisenberg uncertainty relation is modified which can be inferred from a deformed algebra \cite{Kempf, AKempf, Hinr}. This algebra can be generalized to a Lorentz covariant form \cite{Ques, Quesn} and then the commutation relations are written as
\begin{equation}\label{I}
[X^{\mu},P^{\nu}]=-i\hbar \left[(1-\beta P^2)g^{\mu\nu}-2\beta P^{\mu}P^{\nu}\right]\,,
\end{equation}
$$
[X^{\mu},X^{\nu}]=0\,,\,\,\,[P^{\mu},P^{\nu}]=0\,,
$$
where, $P^2=P^{\mu}P_{\mu}=(P^{0})^2-|\vec{P}|^2=\frac{E^2}{c^2}-|\vec{P}|^2$. This generalized algebra is in fact a simplified version of the more general algebra of \cite{Ques} and \cite{Quesn}, where terms of second order in $\beta$ are omitted and the position operators commute \cite{setare}.
Using Eq.(\ref{I}), the relativistic generalized uncertainty relation (RGUP) in $(1+1)$ dimensions becomes
\begin{equation}\label{Ib}
\triangle X \triangle P\geq \frac{\hbar}{2}\left[1+3\beta(\triangle P)^2-\beta\langle(P^0)^2\rangle+\tau \right]\,.
\end{equation}
in which $\tau$ is a function generally independent of $\triangle p$ and $\triangle x$. This relation indicates the existence of a minimal length as $\Delta X_{min}\approx \hbar\sqrt{3\beta}$, where $\beta=\beta_{0}(\frac{l_{p}}{\hbar})^{2}$, $l_{p}$ and $\beta_{0}$ are the Planck length and a dimensionless constant, respectively. The commutation relation Eq.(\ref{I}) can be obtained by the following representation
\begin{equation}\label{Ic}
X^{\mu}=x^{\mu}\,,\,\,\, P^{\mu}=(1-\beta p^2)p^{\mu}\,.
\end{equation}
that $x^{\mu}$ and $p^{\mu}$ are the ordinary position and momentum operators satisfying $[x^{\mu},p^{\nu}]=-i\hbar g^{\mu\nu}$.

Using the RGUP, the consequences of the minimal length hypothesis can be examined in relativistic quantum theory. For example, in \cite{setare} the modification of the Klein-Gordon equation in $3 +1$ dimensions in the presence of RGUP is studied. Also, considering the plane wave solutions, it is shown that the equation naturally describes two massive particles with different masses. This approach defines an upper limit on the modification parameter of the theory, $\beta$. In \cite{Husain}, the effects of RGUP are studied to the propagation of a free scalar field. In \cite{Bosso}, the predictions of RGUP are studied based on the scattering experiment at high energy scales. In this work, using the Ostrogradsky method, the Lagrangian quantum electrodynamics of a complex scalar field are minimally coupled to RGUP and then the corresponding Feynman propagators for the scalar and gauge fields are calculated.

In our previous paper, we have studied the solitary solutions of the nonlinear Klein-Gordon equation in the context of the RGUP. There, to include the spontaneous symmetry breaking, we also considered the $\phi^{3}$ potential besides the $\phi^{4}$ nonlinear self-interacting term. The appearance of the $\phi^{3}$ term is due to the field condensation around the vacuum and spontaneously breaks the symmetry of $\phi \rightarrow -\phi$. Due to the difficulty of the relevant equations, these models are usually investigated via perturbation mechanisms. But, in that work, we have applied a novel method to find the exact and analytical solutions which is known as the $tanh$ or $sech$-method \cite{Malf, Wazwaz1, Malfl}. Using these methods and their modifications \cite{Waz}-\cite{Wakil}, one can be obtained the exact traveling wave solutions for nonlinear equations with solitary, periodic, and rational forms. On the other hand, the importance of solitary waves is that they could be the steady-state counterpart of the Zeldovich pancakes which emerge at nonlinear states of gravitational instability \cite{Zeld1, Zeld2, Kol}. For instance, in the background of Einstein's static universe, some density fluctuations could lead to the final state that is a soliton traveling \cite{Eug}. The Zeldovich pancakes are usually used to understand the cosmic web as a hierarchy of the condensed structures \cite{Zeld1, Zeld2}.

Here, in the framework of (RGUP), we again consider the nonlinear Klein-Gordon field with the potential which contains self-interaction terms $\phi^{4}$ and $\phi^{3}$. Then we apply an extended $\tanh$ method and find some other new solitary solutions. In this path, we assume that the field solutions should be $sech$-functions of segments of space and time. By applying this assumption, several solutions can be found which we select only the normalizable and localized solitary fields. The exact solutions are obtained in two phases - in the presence of the minimal length for one phase and in its absence for the other. The energy spectrum of each field solution is acquired in an analytical form and discussed. Finally, the modification parameter of the corresponding generalized uncertainty principle is estimated.

It should be noticed that the extended $tanh$-method we introduce here is novel and as far as we have searched,  we have not found it in other researches. In our previous work \cite{Mir1}, we have applied the ordinary $tanh$-method  and found some solitary fields. Our previous field solutions are actually the particular solutions of more general solutions we introduce here applying our generalized method.

The organization of this paper is as follows: In the next section the model is presented. In section 3 we introduce our applied generalized $sech$-method. Then, in section 4 we introduce the physical solitary solutions in the framework of the ordinary CFT without considering the minimal length effects. In this case, we find two sets of solutions which are given in sub-sections 4.1 and 4.2. The physical solitary solutions in the presence of the RGUP effects are introduced in section 5. Finally, the conclusions are presented.


\section{The Model}

Let us consider a real relativistic scalar field $\phi(x)$ which represents a chargeless particle with spin $0$. In the framework of the spontaneous symmetry breaking model, the governing equation of this field is obtained from the Lagrangian density

\begin{equation}\label{M1}
{\cal L}_o=\frac{1}{2}\partial_{\nu}\phi\partial^{\nu}\phi-\frac{1}{2}\mu^2\phi^2-\frac{1}{3} v\phi^3-\frac{1}{4} \lambda \phi^4\,,
\end{equation}
where $\mu$ and $\lambda$ are real parameters. This Lagrangian contains the self interaction terms $\phi^4$ and $\phi^3$.  The parameter $v$ is appeared due to the condensation of the field around the vacuum \cite{Kaku}.

In the presence of the minimal length the momentum operators are changed and thus in our model, the covariant derivative is
\begin{equation}\label{M2}
\partial^{\mu}\longrightarrow \left(1+\beta\hbar^2\Box\right)\partial^{\mu}\,,
\end{equation}
where $\Box=\partial_{\nu}\partial^{\nu}$, and the correction term of the first order of $\beta$ is only kept into considerations.
By this definitions, the Lagrangian density Eq.(\ref{M1}) is modified and obtains an additional correction term and can be rewritten as
 \begin{equation}\label{M3}
{\cal L}=\frac{{\mathcal{E}}_P}{(\ell_P)^{D}}\left\{\frac{1}{2}\partial_{\nu}\phi\partial^{\nu}\phi-\frac{\beta_0}{2}~  \Box\phi \Box\phi-\frac{1}{2}\mu^2\phi^2-\frac{1}{3} v \phi^3-\frac{1}{4} \lambda \phi^4\right\}\,.
\end{equation}
Here, $\ell_P$ and ${\mathcal{E}}_P$ are the Planck length and the Planck energy, respectively and $D$ is the dimension of space. In  Eq.(\ref{M3}) the space and time coordinates are dimensionless via the definition
 \begin{equation}\label{M4}
x^{\mu}=(c\,t,\,{\mathbf{x}})\rightarrow \ell_P x^{\mu}=(c\,t_P\,t,\,\ell_P~{\mathbf{x}})\,,
 \end{equation}
with $t_P$ being the Planck time, $t_P=\frac{\ell_P}{c}$. Also, all of the other dimensional quantities are redefined as follows

\begin{equation}\label{M5}
\left\{ \begin{array}{ll}
  \beta_0=2\,\hbar^2\ell_P^{-2}~\beta \,,\,\,\,\,\,\mu\rightarrow\ell_P^{-1}~ \mu \,,\,\,\, \lambda\rightarrow \ell_P^{D-4} {\mathcal{E}}_P^{-1}~   \lambda\,,\\\\
  \phi^2\rightarrow \ell_P^{2-D} {\mathcal{E}}_P~\phi^2\,,\,\,\,\,\,v\rightarrow\ell_P^{\frac{D}{2}-3} {\mathcal{E}}_P^{\frac{-1}{2}}~v\,.
  \end{array}\,
\right.
\end{equation}
The Hamiltonian, in the presence of the minimal length effects, correspondingly becomes

\begin{equation}\label{M6}
{\cal H}=\frac{{\mathcal{E}}_P}{(\ell_P)^{D}}~\partial_0\phi\left(1+\beta_0~\Box\right)\partial^0\phi-{\cal L}\,,
\end{equation}
where the terms up to the first order of $\beta_0$ is considered and ${\cal L}$ is the total Lagrangian given in Eq.(\ref{M3}).
Applying the generalized E\"{u}ler-Lagrange equation , \cite{Barut} to the Lagrangian Eq.(\ref{M3}) leads to the following field equation

\begin{equation}\label{M7}
\Box \phi+\beta_0~\Box\Box\phi+\mu^2 \phi + v \phi^2+\lambda \phi^3=0\,.
\end{equation}
This is the equation that governs the considered scalar field $\phi$ in the presence of the minimal length. This equation has already been solved by using perturbation techniques. However, here and in our previous work \cite{Mir1}, we introduce its exact solitary solutions.

\section{The Hyperbolic Secant Scheme}

Let us now turn to study the traveling waves or the solitary wave solutions of Eq.(\ref{M7}). The solitary waves, in $(1+1)$ dimensions are defined as
\begin{equation}\label{S1}
\phi(x,t)=\psi(X)\,,\,\,\,X=\frac{g}{\gamma}(x-wt)\,,
\end{equation}
where $\psi(X)$ represents the wave.  Here, $X$ is a particular variable which unites the two independent variables $x$ and $t$. Also, $w$ is the velocity of the travelling wave and $\gamma=\sqrt{1-\frac{w^2}{c^2}}$. Also, $\frac{g}{\gamma}$ is a wave factor which its inverse is proportional to the width of the wave $L$, namely $L\propto \frac{\gamma}{g}$.

As a result of Eq.(\ref{S1}), the partial differential equation Eq.(\ref{M7}) converts into the following ordinary differential equation

\begin{equation}\label{S2}
g^4 \beta_0 ~\frac{d^4\psi}{dX^4}-g^2\,\frac{d^2\psi}{dX^2}+\lambda\,\psi^3+v\,\psi^2+\mu^2\,\psi=0\,.
\end{equation}
The solutions $\psi(X)$ are physically acceptable if they are normalizable. This means that the normalization constant $N$ which is defined as

\begin{equation}\label{S5}
N=\frac{\gamma}{g}\int_{-\infty}^{\infty} dX~\left[\psi(X)\right]^2\,
\end{equation}
 should obey the condition $N=1$. We here consider only the localized and bounded solitary waves, namely those obtain  the finite $N$. The solutions with an infinite value of $N$ are omitted due to their nonphysical meaning.  Another important quantity is the energy assigned to the solitary wave which we calculate by the following integration

\begin{equation}\label{S6}
E=\frac{\gamma~\ell_P}{g}\int_{-\infty}^{\infty} dX~{\cal H}(X) \,,
\end{equation}
where ${\cal H}(X)$ is the Hamiltonian density in terms of $X$. By imposing the transformations Eq.(\ref{S1}) into Eq.(\ref{M6}), one obtains

\begin{equation}\label{S7}
{\cal H}=\frac{{\mathcal{E}}_P}{\ell_P}\left\{g^2\left(\frac{w^2}{\gamma^2}+\frac{1}{2}\right)\left(\frac{d\psi}{dX}\right)^2
+\frac{1}{2}\mu^2\psi^2+\frac{1}{3}v\psi^3+\frac{1}{4}\lambda\psi^4-\beta_0~g^4\left[\frac{w^2}{\gamma^2}\frac{d\psi}{dX}\frac{d^3\psi}{dX^3}-
\frac{1}{2}\left(\frac{d^2\psi}{dX^2}\right)^2\right]\right\}\,.
\end{equation}

Here, to find the assuming solitary solutions for Eq.(\ref{S2}), the $tanh$ method of \cite{Wazwaz1} is applied. By this method, one assumes that the solitary solutions can be written in terms of the well-known solitary waves namely the $sech$ or the $tanh$ functions. Let us now consider the $sech$ forms and apply the following transformations

\begin{equation}\label{S8}
z=sech(X)\,,\,\,\,\psi(X)=\Phi(z)\,.
\end{equation}
Under this transformation, all of the derivatives in Eq.(\ref{S2}) get changed and this equation becomes

$$\left\{g^4 \beta_0\left(z^2 -1\right)^{2} z^{4}  \frac{d^{4}}{d z^{4}} +6 g^4 \beta_0\, z^3\left(2 \,z^{4}-3 \,z^{2}+1\right)\frac{d^{3}}{d z^{3}}+z^2\,\left[g^2 \beta_0(36 \,z^{4}-40 z^{2}+7)+z^2-1\right]g^2 \frac{d^{2}}{d z^{2}}\right.$$
\begin{equation}\label{S9}
\left. +z\,\left[g^2 \beta_0 \left(24\,z^{4}-20 z^{2}+1\right) +2\,z^2-1\right] g^2 \frac{d}{d z}+\left(\lambda\,\Phi \! \left(z \right)^{2} +v\,\Phi \! \left(z \right) +\mu^2 \right)\right\} \Phi \! \left(z \right)\,.
\end{equation}

In our previous work \cite{Mir1}, we assumed that the solutions of Eq.(\ref{S9}) can be expressed in $sech$ functions and considered the solutions as

\begin{equation}\label{S10}
\Phi(z)=a_1\,z+ a_2\,z^2\,,
\end{equation}
 where $a_1$ and $a_2$ are two unknown parameters. Then, by inserting Eq.(\ref{S10}) into Eq.(\ref{S9}), one finds an $6$th-degree equation of $z=sech(X)$. To make the equation holds for any $z$,  the coefficients of the terms of any power of $z$ should be zero. In this way, $7$ coupled equations of the unknown parameters of the theory are constructed. Then solving these equations, one can find the appropriate parameters. However, between several obtained solutions, we select only those solutions which can be normalized, namely are localized and find a finite normalization constant $N$. Also, we don't consider a constant term $a_0$ in Eq.(\ref{S10}) which already is in the original works \cite{Wazwaz1}.

However, the ansatz Eq.(\ref{S10}) is not the only expecting answer. Indeed by considering the second-order solutions Eq.(\ref{S10}), the first two terms of Eq.(\ref{S9}) vanish. Therefore solutions Eq.(\ref{S10}) may not be very satisfactory. In fact, this relation displays a polynomial of $z=sech$ and provides the simplest solution of Eq.(\ref{S9}). A more general solution can be represented by the following functional
\begin{equation}\label{S11}
\Phi(z)=\frac{a_0+a_1\,z+ a_2\,z^2}{b_0+b_1\,z+ b_2\,z^2}\,,
\end{equation}
where $a_i$ and $b_i$ are six unknown constants. Here, we assume the ansatz Eq.(\ref{S11}) which provide us a more general solution of Eq.(\ref{S9}) in comparison to Eq.(\ref{S10}).  We should note that Eq.(\ref{S10}) is a special case of Eq.(\ref{S11}), where $a_0$, $b_1$ and $b_2$ are taken to be zero. In this case, by substituting Eq.(\ref{S11}) in Eq.(\ref{S9}), one obtains a $10$th-degree equation of $z$. The coefficient of each power of $z$ must be zero which leads to $11$ coupled equations between parameters of the theory. By solving these equations, one can find the appropriate dealing parameters. In what follows, we turn to investigate these proper sets of solutions and introduce them in two different categories, in the absence ($\beta_0=0$) and in the presence of the RGUP effects ($\beta_0\neq 0$).


\section{The solitary solutions in the ordinary CFT, and $\beta_0=0$}

Without assuming the minimal length scale effects, i.e $\beta_0=0$,  we find two sets of solitary solutions. In fact, one can find several sets of solutions by ascribing the solution in the general form Eq.(\ref{S11}). However, among them, we select the solutions which represent localized and normalizable Klien-Gordon field. With this standard, we find two different sets of solutions. In both sets, we put $b_0=1$, without lose of the generality, and for simplifying the solutions.

\subsection{The first set of solutions}

In the absence of the quantum gravity effects, one set of solutions of the theory are obtained as follows

\begin{equation}\label{F1}
v= -\!\frac{3~g^2}{a_1^2}(a_1~b_1 - a_2)\,,\,\,b_2=\frac{a_2}{a_1^2}~(a_1~b_1 - a_2)\,,\,\,\lambda= \frac{2~g^2}{a_1^4}~(a_1^2~b_1^2 - 2a_1a_2b_1 - a_1^2 + a_2^2)\,,\,\,\mu^2=g^2\,.
\end{equation}
where $a_1$, $a_2$, $b_1$ and $g$ are free parameters. Using these parameters, the solitary wave solution of Eq.(\ref{S11}) becomes

\begin{equation}\label{F2}
\psi(X)=\frac{a_1^{2}~\mathrm{sech}\! \left(X \right) }{a_1+\left( a_1 b_1 -a_2 \right)~\mathrm{sech}\! \left(X \right)}\,.
\end{equation}
This field can be written in a more simpler form

\begin{equation}\label{F3}
\psi(X)=\frac{a_1}{\cosh \! \left(X \right)+u}
\end{equation}
where $u=b_1-\frac{a_2}{a_1}$. In order to fulfil the normalization condition, substituting Eq.(\ref{F3}) into Eq.(\ref{S5}) after integration one finds the normalization constant $N$. Then, by applying the condition $N=1$ the parameter $a_1$ gets
\begin{equation}\label{F4}
a_1^2= \frac{g}{2\gamma}\, \frac{\left(u^{2}-1\right)^{\frac{3}{2}} }{2 u \,\mathrm{arctanh}\! \left(\frac{u -1}{\sqrt{u^{2}-1}}\right)-\sqrt{u^{2}-1}},
\end{equation}
To make the parameter $a_1$ reasonable, we should apply the conditions

 \begin{equation}\label{F5}
\left\{ \begin{array}{ll}
 u\neq \pm 1\\\\
  -1\leq\frac{u-1}{\sqrt{u^2 - 1}}\leq 1
  \end{array}\,.
\right.
\end{equation}
Beside these conditions, the denominator of Eq.(\ref{F4}) should be nonzero and $a_1^2>0$. All of these conditions are satisfied by choosing $u> 1$.  Then by the definition $a_1$ in Eq.(\ref{F4}) and under the condition $u>1$,  the solitary wave Eq.(\ref{F3}) is normalized and well-defined.

The energy of the solitary field can be found by integrating the Hamiltonian density according to Eq.(\ref{S6}). Inserting Eq.(\ref{F3}) into Eq.(\ref{S7}), the Hamiltonian density becomes

\begin{equation}\label{F6}
{\mathcal{H}}=\frac{g^{2} a_1^{2} \sinh^{2}\left(X \right)}{\ell_P \,\gamma^{2} \left( \cosh \! \left(X \right)+u \right)^{4}}{\mathcal{E}}_P\,,
\end{equation}
then by integration one finds

\begin{equation}\label{F7}
E=\frac{g a_1^{2} \left(-6 u \,\mathrm{arctanh}\! \left(\frac{u -1}{\sqrt{u^{2}-1}}\right)+\left(u^{2}+2\right) \sqrt{u^{2}-1}\right)}{3 \left(u^2 -1\right)^{2}\sqrt{u^{2}-1}\, \gamma  }{\mathcal{E}}_P
\end{equation}

\begin{figure}
\begin{center}
\epsfig{file=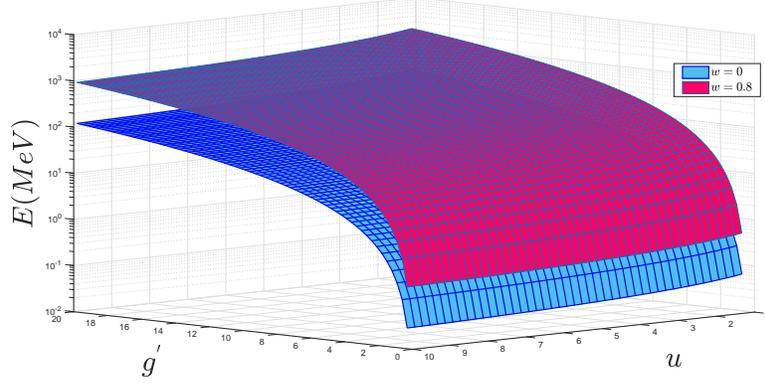,width=12cm}
\caption{\small{The energy spectrum $E$ of the solitary waves corresponding to the masons energy, as a function of $g^{'}$ and $u$. Here, it is considered that $g^{'}=g(1.22\times 10^{22})^{\frac{1}{2}}=1.1 g \times 10^{11} $.}} \end{center}
\end{figure}

In order to examine how this energy may gives correct energy of a meson field, $E$ in Eq.(\ref{F7}) is plotted in Fig.(1).
The rest energy $E$ of a meson field is about $0.51~MeV < E < 940~MeV$ and the Planck energy is ${\mathcal{E}}_P=1.22\times 10^{22}~MeV$. This figure, is plotted for two different values of the velocity parameter, namely $w=0$ and $w=0.8~c$ and illustrates $E$ in $MeV$ respect to $u$ and $g^{'}=g(1.22\times 10^{22})^{\frac{1}{2}}=1.1 g \times 10^{11} $. One should note that $u$ must be in the range $u>1$ to have reasonable localized solutions. As the figure shows the model can imply to describe the energy of a meson field without dealing the complicated traditional methods.

The wave Eq.(\ref{F3}) is  plotted in Fig.(2), with $a_1$ being defined in Eq.(\ref{F4}). Here $\phi$ and $X$ are defined in Eq.(\ref{S1}) and became dimensionless via Eqs.(\ref{M4}) and (\ref{M5}). This figure in fact illustrates  $\frac{\phi^2}{g\ell_P {\mathcal{E}}_P}$ versus $x^{'}=\left(\frac{x}{\ell_P}\right) \frac{g}{\gamma}$ and $t^{'}=\left(\frac{t}{t_P}\right)\frac{g}{\gamma}$. This figure ensures that the wave is a nonsingular  and localized solitary wave.

\begin{figure}
\begin{center}
\epsfig{file=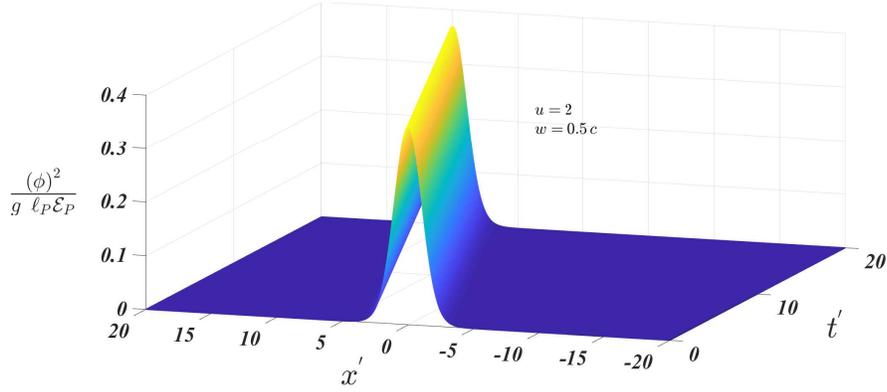,width=12cm}
\caption{\small{The solitary wave $\frac{\phi^2}{g\ell_P {\mathcal{E}}_P}$ as a function of the dimensionless position $x^{'}=\left(\frac{x}{\ell_P}\right) \frac{g}{\gamma}$ and the dimensionless time $t^{'}=\left(\frac{t}{t_P}\right)\frac{g}{\gamma}$.}}\end{center}
\end{figure}

It should be noticed that the solution \ref{F3} is obtained in the presence of the spontaneous symmetry breaking $(v\neq 0)$. In our previous work \cite{Mir1}, we have introduced a solution in the absence of the spontaneous symmetry breaking effects namely, $v=0$ which is a particular case $u=0$ of Eq.(\ref{F3}). Therefore, as we expect, the ansatz Eq.(\ref{S11}) provides us with more general forms of the solitary fields than the simple ansatz Eq.(\ref{S10}).


\subsection{The second set of solutions}

In the case of without considering the minimal length effects, one can find another set of solutions which is specified by

\begin{equation}\label{N1}
b_1= 0\,,\,\,\,a_1= 0\,,\,\,\,v= -\frac{6 g^2}{a_2} \left(2 b_2 + 1\right)\,,\,\,\,\lambda= \frac{8 g^2b_2}{a_2^2}(b_2 + 1)\,,\,\,\,\mu^2= 4g^2\,,\,\,\,
\end{equation}
where $a_2$, $b_2$ and $g$ are free parameters. In this case, the solitary wave solution of Eqs.(\ref{S8}) and (\ref{S11}) gets

\begin{equation}\label{N2}
\psi(X)=\frac{a_2~ \mathrm{sech}\! \left(X \right)^{2}}{b_2~ \mathrm{sech}\! \left(X \right)^{2}+1}\,.
\end{equation}
With more simplifications, the wave field Eq.(\ref{N2}) can be rewritten as
\begin{equation}\label{N3}
\psi(X)=\frac{a_2}{\cosh^{2}\left(X \right)+b_2}\,,
\end{equation}
which is obviously different from the first wave solution Eq.(\ref{F3}). This field solution is obtained in the presence of the spontaneous symmetry breaking, namely for $v\neq 0$, according to Eq.(\ref{N1}). This is a new answer, which can be obtained only via the general ansatz Eq.(\ref{S11}) and can not be found by the ansatz Eq.(\ref{S10}).

In order to study the energy of the field, by inserting Eq.(\ref{N3}) into Eq.(\ref{S7}), the Hamiltonian density this time is obtained as
\begin{equation}\label{N4}
{\mathcal{H}}=\frac{4 a_2^{2}\,g^2\, \cosh^{2}\left(X \right)\sinh^{2}\left(X \right)} {\ell_p \,\gamma^{2}\,\left(\cosh^{2}\left(X \right)+b_2 \right)^{4} }{\mathcal{E}}_P\,.
\end{equation}
Then, by using Eq.(\ref{N4}) into Eq.(\ref{S6}), the energy spectrum can be found. However, by speculation about the relevant localized solutions, it becomes obvious that the solutions of this case must be separated into two different subcases with different behaviors.

Hence, before turning to derive the relevant energy spectrum corresponding to Eq.(\ref{N4}), let us now turn to study the other important key point, namely the normalization of the field which in turn characterizes the localizable and bounded solitary solutions. Substituting Eq.(\ref{N3}) into Eq.(\ref{S5}), the normalization constant $N$ can be obtained. More speculations in this regard show that the field solutions are not bounded  for $b_2\leq -1$. In fact since $cosh(X)\geq 1$, the denominator of Eq.(\ref{N3}) have two poles at $X=\pm \sqrt{-b_2}$ for $b_2\leq -1$. Therefore, the field solutions Eq.(\ref{N3}) for $-1<b_2$ are bounded,  normalizable and $N$ can be found in a closed analytic form.

 On the other hand, more speculations show that the regime $-1<b_2$ also should be divided into two different regimes $-1<b_2<0$ and $0<b_2$. In fact, we find that the integrations Eqs.(\ref{S5}) and (\ref{S6}) give different results in the intervals $-1<b_2<0$ and $0<b_2$. Two different functions for the normalization constant $N$, and for the energy can be found in these two intervals. So let us now break this section into the following two subsections.


\subsubsection{$-1<b_2<0$}

Putting Eq.(\ref{N3}) into Eq.(\ref{S5}), for the interval $-1<b_2<0$, one finds $N$ in terms of $b_2$, $g$ and $\gamma$. Solving the consequent relation $N=1$,   the parameter $a_2$ finds the analytic formula

\begin{equation}\label{A1}
a_2 =\pm\left\{\frac{g}{\gamma}\,\frac{-b_2~ \left(1-\sqrt{-b_2} \right) \left(1+b_2 \right)^{\frac{3}{2}}}{ 2 (1+2b_2)\arctan \! \left(\frac{\sqrt{b_2 +1}}{1-\sqrt{-b_2}}\right) +\sqrt{1+b_2}\,(1-b_2 -2 \sqrt{-b_2})^{\frac{1}{2}}}\right\}^{\frac{1}{2}}\,.
\end{equation}
It should be noticed that this parameter $a_2$ is not real valued throughout of the interval $-1<b_2<0$. In fact the denominator of Eq.(\ref{A1}) has a root at $b_2\approx -0.54$ in the interval $-1<b_2<0$ and is negative for $-1<b_2<-0.54$. This fact is illustrated in Fig.(3), where $\frac{\gamma\,a_2^2}{g}$ is plotted against $b_2$. According to the figure, at $b_2=-0.54$, $a_2^2$ changes sign and for $-1<b_2<-0.54$, $a_2^2<0$. Therefore, $a_2$ in Eq.(\ref{A1}) is real valued in $-0.54<b_2<0$.

\begin{figure}\begin{center}
\epsfig{file=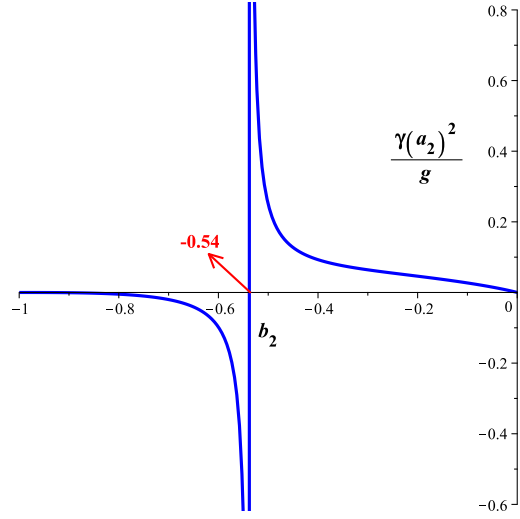,width=7cm}
\caption{\small{The parameter $\gamma\frac{a_2^2}{g}$ of Eq.(\ref{A1}) as a function of $b_2$. }}\end{center}
\end{figure}

In Figs.(4-a) and (4-b), the normalized amplitude $\frac{\gamma\psi^2}{g}$ is plotted against $X$ and $b_2$ where $a_2$ is given in Eq.(\ref{A1}) for $-0.54<b_2<0$. In Fig.(4-a) the interval $-0.54<b_2<-0.3$ and in Fig.(4-b) the interval $-0.3<b_2<0$ are considered. As the figure shows, by approaching $b_2 \rightarrow ^{-}0$ the wave becomes wider and wider.

\begin{figure}\begin{center}
\epsfig{file=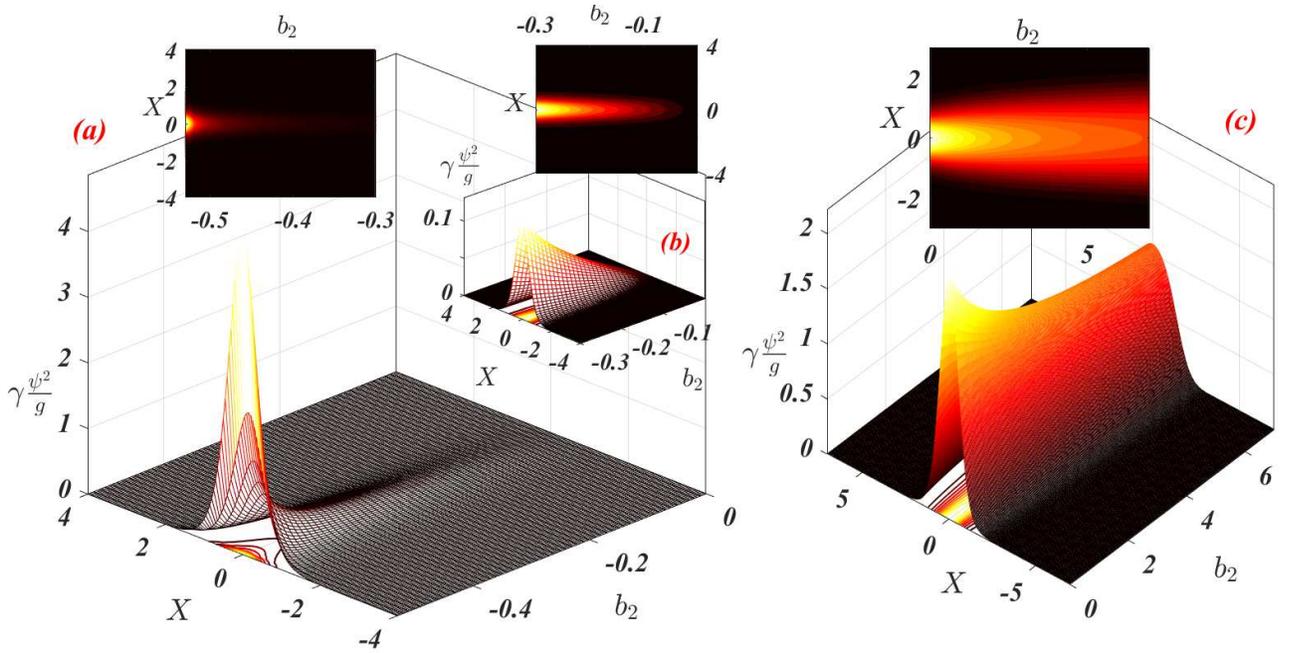,width=17cm}
\caption{\small{The normalized field $\frac{\gamma\psi^2}{g}$ in Eq.(\ref{N3}) of the second set of solutions, section $4.2$, versus the position $X$ and the parameter $b_2$. The parameter $b_2$ is settled in different range of $-0.54<b_2$, namely in $(a)$ $-0.54<b_2<-0.3$, in (b) $-0.3<b_2<0$ and in $(c)$ $0<b_2$. In Figs.(a) and (b), the parameter $a_2$ is given in Eq.(\ref{A1}) and in Fig.(c), it is given in Eq.(\ref{E1}).}}\end{center}
\end{figure}

The energy of the solitary wave, obtained in this subsection, can be found by integrating the Hamiltonian density Eq.(\ref{N4}) in according to Eq.(\ref{S6}). For the interval $-0.54<b_2<0$ this integration leads to

\begin{equation}\label{A2}
E=\frac{2 g\, \left\{3(\frac{1-\sqrt{-b_2}}{\sqrt{b_2 +1}})\left(b_2+ \frac{1}{2}\right) \arctan \! \left(\frac{\sqrt{b_2 +1}}{1-\sqrt{-b_2}}\right)- \left[b_2^{3}-\frac{b_2}{4}+\sqrt{-b_2}\left(\frac{3}{2}-2b_2-2b_2^2\right)-\frac{3}{4}\right]\right\}\, a_2^{2}}{3 \left(b_2 +1\right)^{2} \gamma \left(1-\sqrt{-b_2}\right)^2\,b_2^{2}}\,
{\mathcal{E}}_P\,.
\end{equation}
The denominator of this energy relation is always positive. However the numerator of this energy relation change sign at $b_2\approx -0.3$. Therefore for $-0.54<b_2<-0.3$  the energy is negative, but for $-0.3<b_2<0$ the energy is positive. This fact is illustrated in Fig.(5) where the energy $\frac{E}{g\, {\mathcal{E}}_P}$ of Eq.(\ref{A2}) is represented as a function of $b_2$. Fig.(5-a) contains the energy spectrum corresponding to the interval $-0.54<b_2<-0.3$, but Fig.(5-b) illustrates the energy for $-0.3<b_2$ which also contains the energy of the following subsection namely $b_2>0$. As Fig.(5-a) obviously shows, the energy is negative in the interval $-0.54<b_2<-0.3$ which indicates the presence of the bound state solutions. However, for the other part of the interval which is represented in Fig.(5-b), namely for $-0.3<b_2<0$ the energy is positive.

\begin{figure}\begin{center}
\epsfig{file=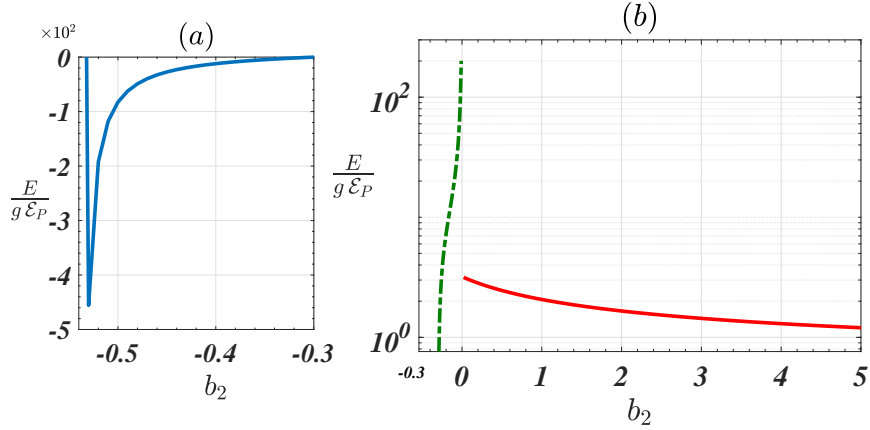,width=12cm}
\caption{\small{The energy $\frac{E}{g\,{\mathcal{E}}_P}$ as a function of $b_2$ for the full interval $-0.54<b_2$. Fig.(a) illustrates the energy spectrum for $-0.54<b_2<-0.3$ according to Eq.(\ref{A2}). The Fig.(b) shows the energy for the range $-0.3<b_2$, where for $-0.3<b_2<0$ the relation Eq.(\ref{A2}), and for $0<b_2$ the relation Eq.(\ref{E2}) are used.}}\end{center}
\end{figure}

\subsubsection{$0<b_2$}
For the interval $0<b_2$, the normalization constant $N$ obtains another relation from the relation of the previous subsection. Then by putting the obtained $N$ into the normalization condition $N=1$, one finds a set of relations for the parameter $a_2$ as follows
\begin{equation}\label{E1}
a_2 =\pm\left\{
\frac{2 ~ g~ \left[b_2(1+b_2) \right]^{\frac{3}{2}}}{\gamma \left[(2 b_2+1)\ln \left(\frac{\sqrt{1+b2}+\sqrt{b2}}{-\sqrt{b2}+\sqrt{1+b2}}\right) -2 \sqrt{b_2(1+b_2)}\right]}\right\}^{\frac{1}{2}}
\end{equation}
The normalized field amplitude $\frac{\gamma\psi^2}{g}$ in Eq.(\ref{N3}) with $a_2$ of Eq.(\ref{E1}) is plotted in Fig.(4-c). As the figure shows, in this range of $0<b_2$, the field is localized and normalizable and well- behaved. Also Fig.(4.c) indicates that by growing the parameter $b_2$ the hight of the solitary wave becomes slightly shorter and its width becomes slightly wider.

The spectrum of energy of this case can be found by inserting Eq.(\ref{E1}) into Eq.(\ref{N4}), and then integrating in according to Eq.(\ref{S6}) which leads to the following analytical solution

\begin{equation}\label{E2}
E=\frac{\left[\left(\frac{1}{2}+b_2 \right) \ln \left(\frac{-\sqrt{b_2}+\sqrt{b_2 +1}}{\sqrt{b_2}+\sqrt{b_2 +1}}\right) +\sqrt{b_2(b_2 +1)}\, \left(1+\frac{4 }{3}b_2+\frac{4 }{3}b_2^2\right)\right] a_2^{2}~ g }{2 b_2^{\frac{5}{2}} \left(b_2 +1\right)^{\frac{5}{2}} \gamma}{\mathcal{E}}_P\,.
\end{equation}
This energy is plotted in Fig.(5-b) versus the parameter $b_2$. In fact, the $0<b_2$ part of Fig.(5-b) illustrates the energy spectrum of Eq.(\ref{E2}) where the parameter $a_2$ in Eq.(\ref{E1}) is applied. As the figure shows in the interval $0<b_2$ the energy is positive and very slightly reduces by increasing the value of the parameter $b_2$.


\section{The solitary solutions in the RGUP model, for $\beta_0\neq 0$}

Let us now turn to study the solitary field solutions by the supposition of the presence of the minimal length, namely when $\beta_0\neq 0$. In this case, the equations contain one more parameter, in comparison  with the previous section.  We make the assumptions $a_0=0$, $ b_0=1$ and $b_1=0$ to make the situation more simpler. By these assumptions, we find the following set of solutions.

\begin{equation}\label{O1}
a_1= -a_2\,,\,\,\,b_2= -1\,,\,\,\,v= -\frac{3g^2}{a_2}(5\beta_0\,g^2 - 1)\,,\,\,\lambda= -\frac{30\beta_0}{a_2^2}\,g^4\,,\,\,\,\mu^2= -g^2(\beta_0\,g^2 - 1)\,,
\end{equation}
where $a_2$, $\beta_0$ and $g$ are arbitrary parameters. Substituting Eq.(\ref{O1}) into Eqs.(\ref{S8}) and (\ref{S11}), the solitary wave becomes
\begin{equation}\label{O2}
\psi(X)=-\frac{a_2 \,\mathrm{sech}\! \left(X \right) }{\mathrm{sech}\! \left(X \right)+1}\,,
\end{equation}
which by more simplifications, can be rewritten as

\begin{equation}\label{O3}
\psi(X)=-\frac{a_2}{1+\cosh \! \left(X \right)}\,.
\end{equation}
This solution is similar to the first set of solutions in the previous section, namely Eq.(\ref{F3}) for $u=1$. However $u=1$ is not in the accepted interval of Eq.(\ref{F5}). Therefore Eq.(\ref{O3}) is a different solution that is obtained in the presence of the minimal length effects.
Now to normalize the field, we note that by substituting Eq.(\ref{O3}) into Eq.(\ref{S5}) and after integration one finds

\begin{equation}\label{O4}
N= \frac{2\,\gamma\, a_2^2}{3\,g}\,.
\end{equation}
Then, by applying the normalization condition $N=1$ the parameter $a_2$ is obtained

\begin{equation}\label{O5}
a_2=\pm \frac{\sqrt{6\,\gamma\,g}}{2\,\gamma}\,,
\end{equation}
which ensures the normalization of the solitary wave Eq.(\ref{O3}).

Now, by inserting Eq.(\ref{O3}) into Eq.(\ref{S7}), the Hamiltonian density becomes
\begin{equation}\label{O6}
{\mathcal{H}}=\frac{\,g^{2} a_2^{2} \left\{\left[1+\beta_0 \left(\gamma^{2}-1\right) g^{2}\right] \cosh^{2}\left(X \right)-4 g^{2} \beta_0\left[ \left(\gamma^{2}-\frac{3}{2}\right) \cosh \! \left(X \right)-4 \gamma^{2}+5\right] -1\right\}}{\ell_P \,\gamma^{2} \left(\cosh \! \left(X \right)+1\right)^{4}}\,{\mathcal{E}}_P\,.
\end{equation}
The energy of the solitary wave is obtained by integrating the Hamiltonian density Eq.(\ref{O6}), in according to Eq.(\ref{S6}), and one finds

\begin{equation}\label{O7}
E=\frac{\left[5 \beta_0 \,g^{2} (\gamma^{2}+1) +7\right] g^{2} }{35 \gamma^{2}}\,{\mathcal{E}}_P\,.
\end{equation}

On the other hand, another set of solutions in the case of $\beta_0\neq 0$ is given by the following relations of the parameters
\begin{equation}\label{H1}
a_1= 0\,,\,\,\,b_2= 0\,,\,\,\,v= \frac{6\,g^2}{a_2}\,(20\,\beta_0\,g^2 - 1)\,,\,\,\,\lambda= -\frac{120\beta_0\,g^4}{a_2^2}\,,\,\,\mu^2= -16\,\beta_0\,g^4 + 4\,g^2\,,
\end{equation}
where $a_2$, $\beta_0$  and $g$ are arbitrary parameters. Inserting Eqs.(\ref{H1}) into Eqs.(\ref{S8}) and (\ref{S11}), the corresponding field becomes

\begin{equation}\label{H2}
\psi(X)=a_2\,\mathrm{sech}^{2}\left(X\right)\,,
\end{equation}
This field function has been also obtained in our previous article \cite{Mir1}. It should be mentioned that the set Eq.(\ref{H1}) can transform to the set Eq.(\ref{O1}), by replacing

\begin{equation}\label{H3}
a_2\rightarrow -2a_2\,,\,\,\,g\rightarrow 2g\,,
\end{equation}
in Eq.(\ref{O1}). In this way, the main parameters $\lambda$, $\mu^2$ and $v$  in these two sets become equal. Also, by considering the equality $\mathrm{sech}^{2}\left(y\right)=\frac{1}{1+\cosh\left(2y \right)}$
and the replacement Eq.(\ref{H3}), the two solitary fields Eqs.(\ref{O3}) and (\ref{H2}) becomes similar.
However, in this latter case, by applying the normalization condition, one finds

\begin{equation}\label{H4}
 a_2=\pm\frac{\sqrt{3\gamma~g}}{2\gamma}\,.
\end{equation}
Also, the energy becomes
\begin{equation}\label{H5}
E={\frac { 4\left[ 20\,{\it \beta_0}\,{g}^{2}({\gamma}^{2}+1)+
7\right] g^2}{35\,\gamma^2}{\mathcal{E}}_P}\,.
\end{equation}

Let us now turn to study the energy in Eq.(\ref{O7}) and  in the static situation where $\gamma=1$. The energy equation Eq.(\ref{H5}) can equally be studied  which obtains no different conclusions. As a physical restriction, we first note that $\mu^2>0$ to have real $\mu$. To satisfy this situation and also obtain positive energy, we recognize the two following different cases which can be deduced from Eqs.(\ref{O1}) and (\ref{O7})

\begin{enumerate}
  \item For $\beta_0<0$\,,\,\,\,  $-0.7<\beta_0 g^2<0$;
  \item For $\beta_0>0$\,,\,\,\,\,\,\,\,\,\,\,\,\,  $0<\beta_0 g^2<1$\,.
\end{enumerate}
These two situations can be unified as $-0.7<\beta_0 g^2<1$. The corresponding energy diagram is plotted in Fig.(6). The figure is showing the rest energy ($w=0$) as a function of $g^{'}$ and $\beta_0^{'}$  which are defined as

\begin{equation}\label{H6}
g^{'}=g\,\sqrt{1.22\times10^{22}}\,,\,\,\,\beta_0^{'}=\frac{\beta_0}{1.22\times 10^{22}}\,.
\end{equation}
These parameters are considered in such a way to give the energy in $MeV$. The energy $E$ is given in Eq.(\ref{O7}) and is supposed to be suitable to describe the energy of a typical meson field.

\begin{figure}\begin{center}
\epsfig{file=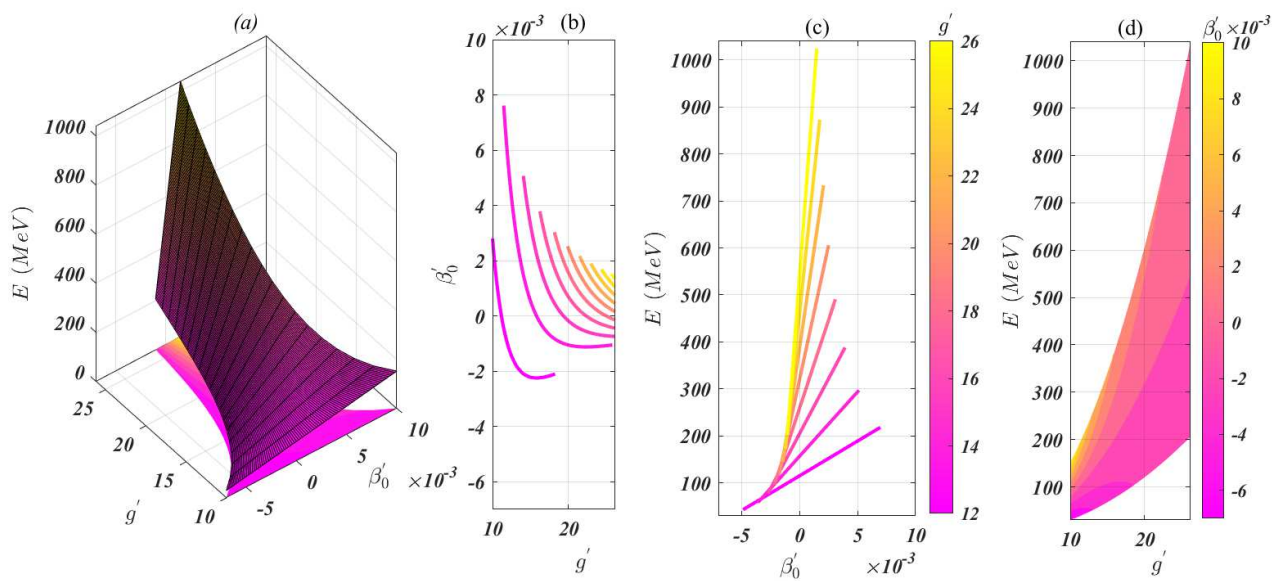,width=17cm}
\caption{\small{The surface (a) shows the energy $E$ as a function of $g^{'}$ and $\beta_0^{'}$. The contour diagram (b) plots $\beta_0^{'}$ respect to  $g^{'}$ , the contour diagram (c) shows $E$ verses $\beta_0^{'}$ (b), and the contour diagram (d) plots $E$ versus $g^{'}$.}}\end{center}
\end{figure}

Fig.(6) indicates that by growing $g^{'}$ the rest energy increases. The parameter $g^{'}$ is proportional to the inverse of the width of the corresponding solitary wave. Hence, the figures show that narrower waves carry higher energies. But the behavior of the energy diagram concerning the parameter $\beta_0^{'}$  shows that heavier mesons, the mesons with higher energies, correspond to  smaller values of $\beta_0^{'}$.  This latter fact can be seen better in Fig.(6~c), where the diagram concentrated around smaller values of $\beta_0^{'}$. These values of $\beta_0^{'}$ cover all the energy from $100~ MeV$ to $1000~ MeV$. However, by increasing the magnitude $|\beta_0^{'}|$, the energy decreases. On the other hand, the minimum scale of length $\Delta x_{min}$ is proportional to the square root of $|\beta_0|$, so the model predicts that the minimal length in the presence of the heavier meson fields can be smaller than that the minimal length in the presence of the lighter meson fields.

Let us now turn to study $\beta_0$ which is the modification parameter of the RGUP model that here we incorporate into the standard spontaneous symmetry breaking theory. This parameter is important because its order of magnitude in fact specifies the possible minimal length scale via the relation $\Delta x_{min}\approx \sqrt{\frac{3}{2}\beta_0}\ell_P$. The model we consider here provides us a way to find an approximation of the parameter $\beta_0$. To do this, we estimate the minimal length by the width of the solitary wave corresponding to the heaviest meson field ever seen, namely the Higgs boson. The width of the meson field decreases with the increase of energy. So, we expect that the width of the heaviest meson is the smallest value of the length that can be observed. Here, we find the width of the solitary wave by using the points corresponding to the half of the height of the maximum of $\psi^2_m$. The height of the peak of the field $\psi^2$  in Eq.(\ref{O3}) is given by $\psi^2_m=\psi^2(X=0)=\frac{3g}{8}$. Then by solving the equation $\psi^2-\frac{1}{2}\psi^2_m$ for $X$ we obtain the acceptable value

\begin{equation}\label{H7}
X=arccosh(-1 + 2\sqrt{2}) \,
\end{equation}
which gives the width of the wave.
Then by using Eqs.(\ref{S1}) and (\ref{M4}), the width of the field $\sigma$ becomes
\begin{equation}\label{H8}
\sigma=\frac{arccosh(-1 + 2\sqrt{2})}{g}\ell_P\approx \frac{1.21}{g}\ell_P\,.
\end{equation}
Now, we assume that the promised minimal length $\Delta x_{min}\approx \sqrt{\frac{3}{2}\beta_0}\ell_P$ is proportional to the width Eq.(\ref{H8}), namely $\Delta x_{min}=\sigma$. In this way, we find

\begin{equation}\label{H9}
\beta_0=\frac{0.98}{g^2}\,.
\end{equation}
To estimate $\beta_0$, we assume that the minimal length should be correspond to the heaviest meson field, namely the Higgs boson, with the rest mass $E=125.35~GeV$. Inserting Eq.(\ref{H9}) into Eq.(\ref{O7}) and constructing the equation $E=125.35~GeV$, and then solving the equation for $g^{'}$, one finds

\begin{equation}\label{H10}
g^{'}=511.20\,.
\end{equation}
Then, substituting Eq.(\ref{H10}) into Eq.(\ref{H9}) gets $\beta_0^{'}=3.7\times 10^{-6}$. Hence, from Eq.(\ref{H6}) one finds

\begin{equation}\label{H11}
\beta_0\approx 4.74\times 10^{16}\,.
\end{equation}
This is an interesting estimation of the modification parameter of the theory. This estimation is obtained by using the Higgs masses. In fact considering a boson field with higher rest energy, leads to smaller values of $\beta_0^{'}$.

The estimation Eq.(\ref{H11}) can be compared to the ones predicted by different approaches. This estimation is close to the predicted value by the electroweak scale and the potential barrier problem, but it is more stringent than that obtained by the Lamb shift effect \cite{Das1,Das2}. Also, this estimation is far from the upper bound obtained for some lower intermediate energies such as the corrections of the Morse spectrum \cite{Mir2} or for the accurate thermodynamics measurements \cite{Mir3}-\cite{Mir5}.

\section{Conclusions}
In this paper, we considered a real relativistic scalar field $\phi(x)$ which represents a chargeless particle with spin 0. Assuming the existence of the minimal length, in the framework of the spontaneous symmetry breaking model, we obtained the governing equation of the field and studied its traveling waves or solitary wave solutions. Here, to find the solitary exact solutions, we applied an extended tanh approach in which the solitary solutions can be written in terms of the well-known solitary waves namely the sech or the tanh functions. It should be noted, that between several obtained exact solutions, we selected only those solutions which could be normalized. In this work, we assumed a more comprehensive initial ansatz which provided us the more general solutions in comparison to the ones of \cite{Mir1}. In this way, a 10th-degree equation was obtained which was divided into 11 coupled relations. By solving these relations, we could obtain the parameters of the theory. The exact solutions were obtained in two phases - in the CFT theory ($\beta_{0} = 0$) and in the presence of RGUP effects ($\beta_{0} \neq 0$). The energies of the solitary fields were found by integrating the Hamiltonian density and to examine how these energies may give the correct energy of a meson field, the energies are plotted. These figures show the model can imply to describe the energy of a meson field without dealing with the complicated traditional methods. These figures ensure that the waves are the nonsingular localized solitary waves. We also observed that the modification parameter of the considered RGUP model depends on the rest energy of the meson fields. Considering the mass of the Higgs boson, we have estimated the modified parameter of the theory.


\end{document}